\documentclass[longauth]{aa}  

\usepackage{graphicx}
\usepackage{graphics,epsfig,color}
\usepackage{natbib}
\usepackage{amsmath}
\bibpunct{(}{)}{;}{a}{}{,}

\usepackage[varg]{txfonts}
\newcommand{\ergcm}[1]{$\times 10^{#1}$ erg cm$^{-2}$ s$^{-1}$}

\newcommand{\hcm}[1]{$\times 10^{#1}$ cm$^{-2}$}

\newcommand{\nh}{$N_\mathrm{H}$}

\newcommand{\ltsima}{$\buildrel < \over \sim$}
\newcommand{\lsim}{\lower.5ex\hbox{\ltsima}}
\newcommand{\gtsima}{$\buildrel > \over \sim$}
\newcommand{\gsim}{\lower.5ex\hbox{\gtsima}}

\newcommand{\xmm}{\emph{XMM-Newton}}
\newcommand{\hess}{H.E.S.S.}

\newcommand{\chandra}{\emph{Chandra}}

\newcommand{\g}{G\,15.4$+$0.1}
\newcommand{\hessj}{HESS\,J1818$-$154}
\begin{document}


%

\title{\hessj, a new composite supernova remnant discovered in TeV gamma rays and X-rays}
\authorrunning{H.E.S.S. Collaboration}
\titlerunning{\hessj}
\author{H.E.S.S. Collaboration
\and A.~Abramowski \inst{1}
\and F.~Aharonian \inst{2,3,4}
\and F.~Ait Benkhali \inst{2}
\and A.G.~Akhperjanian \inst{5,4}
\and E.~Ang\"uner \inst{6}
\and G.~Anton \inst{7}
\and S.~Balenderan \inst{8}
\and A.~Balzer \inst{9,10}
\and A.~Barnacka \inst{11}
% ...
}
  \offprints{\\P. Hofverberg \email{petter.hofverberg@mpi-hd.mpg.de};\\P. Eger, \email{peter.eger@mpi-hd.mpg.de}}

\institute{
Universit\"at Hamburg, Institut f\"ur Experimentalphysik, Luruper Chaussee 149, D 22761 Hamburg, Germany \and
Max-Planck-Institut f\"ur Kernphysik, P.O. Box 103980, D 69029 Heidelberg, Germany \and
Dublin Institute for Advanced Studies, 31 Fitzwilliam Place, Dublin 2, Ireland \and
National Academy of Sciences of the Republic of Armenia, Yerevan  \and
Yerevan Physics Institute, 2 Alikhanian Brothers St., 375036 Yerevan, Armenia \and
Institut f\"ur Physik, Humboldt-Universit\"at zu Berlin, Newtonstr. 15, D 12489 Berlin, Germany \and
Universit\"at Erlangen-N\"urnberg, Physikalisches Institut, Erwin-Rommel-Str. 1, D 91058 Erlangen, Germany \and
University of Durham, Department of Physics, South Road, Durham DH1 3LE, U.K. \and
DESY, D-15738 Zeuthen, Germany \and
Institut f\"ur Physik und Astronomie, Universit\"at Potsdam,  Karl-Liebknecht-Strasse 24/25, D 14476 Potsdam, Germany \and
Nicolaus Copernicus Astronomical Center, ul. Bartycka 18, 00-716 Warsaw, Poland
% ...
}

\date{Received 25 October 2013 / Accepted 9 December 2013}

   \abstract{Composite supernova remnants (SNRs) constitute a small subclass of the remnants of massive stellar explosions where non-thermal radiation is observed from both the expanding shell-like shock front and from a pulsar wind nebula (PWN) located inside of the SNR. These systems represent a unique evolutionary phase of SNRs where observations in the radio, X-ray, and $\gamma$-ray regimes allow the study of the co-evolution of both these energetic phenomena. In this article, we report results from observations of the shell-type SNR \g\ performed with the High Energy Stereoscopic System (H.E.S.S.) and \xmm . A compact TeV $\gamma$-ray source, \hessj, located in the center and contained within the shell of \g\ is detected by \hess\ and featurs a spectrum best represented by a power-law model with a spectral index of $-2.3 \pm 0.3_{stat} \pm 0.2_{sys}$ and an integral flux of F$(>$0.42$\,\mathrm{TeV}$)=($0.9 \pm 0.3_{\mathrm{stat}} \pm 0.2_{\mathrm{sys}}) \times 10^{-12}$\,cm$^{-2}$\,s$^{-1}$. Furthermore, a recent observation with \xmm\ reveals extended X-ray emission strongly peaked in the center of \g. The X-ray source shows indications of an energy-dependent morphology featuring a compact core at energies above 4\,keV and more extended emission that fills the entire region within the SNR at lower energies. 
Together, the X-ray and VHE $\gamma$-ray emission provide strong evidence of a PWN located inside the shell of \g\, and this SNR can therefore be classified as a \emph{composite} based on these observations. The radio, X-ray, and $\gamma$-ray emission from the PWN is compatible with a one-zone leptonic model that requires a low average magnetic field inside the emission region. An unambiguous counterpart to the putative pulsar, which is thought to power the PWN, has been detected neither in radio nor in X-ray observations of \g.
   }

   
   \maketitle
%

\section{Introduction}
\label{sec-intro}
In the aftermath of supernova explosions, a rich variety of highly energetic non-thermal processes can be observed. At the shock front of the expanding supernova remnant shell where the ejecta from the explosion interacts with the surrounding interstellar matter, charged particles can be accelerated up to energies of TeV, and possibly PeV, through the standard diffusive shock mechanism (DSA) \citep[for a review, see][]{1997vhep.conf...97B} and can give rise to intense broad-band electromagnetic radiation. The non-thermal radiation is produced either by energetic hadrons through production and subsequent decay of neutral pions, or by leptons that radiate through synchrotron and inverse Compton (IC) emission.

The supernova explosion can also leave behind a rapidly rotating neutron star that constitutes a second major source of non-thermal emission. This can take the form of pulsed emission from the pulsar's magnetosphere and/or of an extended pulsar wind nebula (PWN) powered by a relativistic particle outflow from the central engine \citep[for a comprehensive review see][]{2006ARA&A..44...17G}.

The particle wind from the pulsar is decelerated by the inertia of the higher density outer medium, and a termination shock is formed. In the wind, particles are advected outward toward the termination shock, following the frozen-in magnetic field lines in an orderly flow, resulting in a (synchrotron) under-luminous feature close to the pulsar. At the shock front, the particles are isotropized and accelerated, and subsequently produce both synchrotron and IC radiation when gyrating in the nebular magnetic field in the downstream flow. 

Typical broad-band spectral energy distributions (SEDs) of PWNe appear to be dominated by leptonic processes and show two broad peaks. The first one extends from radio to X-ray energies arising from synchrotron radiation, whereas the second one is peaked in the $\gamma$-ray regime that arises from IC up-scattering of ambient low-energy photon fields. 

X-ray observations of PWNe with high angular resolution instruments, such as \xmm , are particularly useful since they may differentiate between the extended synchrotron emission from the PWN, seen downstream of the termination shock, and the magnetospheric non-thermal emission from the pulsar itself \citep[for a review of recent results obtained in X-rays, see][]{2012AIPC.1505..177U}. By comparing the properties of these two components, one can learn about the physical processes responsible for the efficient particle acceleration mechanism energizing PWNe. Very-high-energy (VHE, $E>100$\,GeV) $\gamma$-ray observations of PWNe are complementary to X-ray observations since they trace particles with a longer lifetime, through an emission mechanism largely independent of the nebular magnetic field, unlike the synchrotron X-ray emission. 

X-ray and VHE $\gamma$-ray observations have established two classes of PWNe based on their observational properties. In \emph{young systems}, such as the Crab nebula, the pulsar is still close to its birthplace at the center of the SNR and surrounded by an X-ray and a VHE $\gamma$-ray nebula, and the latter is normally unresolved owing to its small size and the relatively poor angular resolution of the ground-based instruments. In \emph{middle aged systems}, such as HESS\,J1303$-$631 \citep{2012A&A...548A..46H} or HESS\,J1825$-$137 \citep{2006A&A...460..365A}, the IC nebula is found to be significantly extended and offset from the pulsar, in contrast to the much smaller and still pulsar-centered X-ray nebulae. This difference in size is generally attributed to the much shorter cooling timescale of the highest energy lepton population that produces X-ray synchrotron radiation, compared to the lower energy leptons giving rise to the IC component. 
Additionally, for these older systems, the interaction between the expanding PWN and the reverse shock of the SNR likely leads to a compression of the PWN followed by a re-expansion phase. If the SNR evolves into an inhomogeneous interstellar medium (ISM), the impact of the reverse shock on the PWN evolution may vary within the nebula, which may lead to the asymmetric and offset emission regions that are particularly visible in the IC regime. For a detailed study of evolved PWNe in SNRs, see \citet{2001ApJ...563..806B}.

A rare subtype of SNRs/PWNe include so-called \emph{composite} SNRs \citep{1987ApJ...314..203H}, which show distinct evidence of both a PWN and a SNR shell \citep[like G\,0.9$+$0.1 and G\,21.5$-$0.9, see e.g.,][]{2005A&A...432L..25A,2005AdSpR..35.1099M}. 
These systems are excellent laboratories for investigating the co-evolution of PWNe and SNRs and, particularly, the potential interaction between the two. To explore the general nature of particle acceleration, transport, and energy-loss mechanisms in such environments, it is necessary to increase the number of well-studied systems in various stages of evolution. For this purpose, detailed multiwavelength studies have to be performed to define the properties and environment of the SNR, as well as to probe the full extent of the non-thermal spectral energy distribution of the PWN.

The SNR \g\ is a poorly studied object that was initially discovered in a 90\,cm survey of the inner Galaxy conducted by the Very Large Array (VLA) \citep{1538-4357-639-1-L25}. The SNR was reported to have a shell-like morphology with a size of about $14\arcmin\times15\arcmin$ and an average spectral index $\alpha = -0.6 \pm 0.2$, indicating that the radio emission is dominated by non-thermal synchrotron emission from the shell. A recent study by \citet{castelletti2013} places the SNR at a distance of ($4.8\pm1.0$)\,kpc and discovered a molecular cloud coincident with the SNR.
For the reasons outlined above, this source is a potential emitter of both X-rays and VHE $\gamma$ rays. Interestingly, this SNR is slightly larger than the \hess\ point-spread-function (PSF; $\sim$6\arcmin), which allows a morphological comparison of the VHE $\gamma$-ray emission with the size of the SNR shell. 

In this paper the nature of the newly detected VHE $\gamma$-ray source \hessj\ coincident with SNR \g\ is investigated, based on results obtained with H.E.S.S.,\ as well as from a recent X-ray observation performed with \xmm . The morphology and spectral energy distribution are discussed in the context of a composite SNR scenario.

\section{H.E.S.S. observations and results}
\label{sec-tev}

\subsection{The H.E.S.S. Telescope Array}
The High Energy Stereoscopic System \citep[\hess][]{2006A&A...457..899A} is an array of imaging atmospheric Cherenkov telescopes for VHE $\gamma$-ray astronomy that detects Cherenkov light emitted from $\gamma$-ray- induced air showers. \hess\ is comprised of four identical 12\,m diameter telescopes in a square of 120\,m and is located on the southern hemisphere in the Khomas highland of Namibia (latitude $23\degr16\arcmin17\arcsec S$) at a height of 1800\,m above sea level. Employing a stereoscopic technique for detecting air showers, an angular resolution better than $0.1^{\circ}$, and an energy resolution of 15\% are achieved, coupled with a high background rejection power. Its unprecedented sensitivity to $\gamma$ rays permits \hess\ to detect a point source with a flux of 1\% of the Crab nebula at a significance of 5\,$\sigma$ in about 25\,h of observations. In 2012, the \hess\ observatory was extended with a fifth $\sim$28\,m diameter telescope at the center of the array - H.E.S.S. II, which will improve the sensitivity of the instrument, particularly in the energy regime above a few tens of GeV. 

\subsection{Data and analysis methods}
The region around \g\ was observed with H.E.S.S. between 2004 and 2011; the dataset consists primarily of observations from the H.E.S.S. Galactic Plane Survey \citep{2006Natur.439..695A} and offset observations of nearby sources. The dataset in this region has a live time of $\sim$120\,h after standard H.E.S.S. quality selection \citep{2006A&A...457..899A}, although the effective live time is considerably lower ($\sim$80\,h) because of the large average offset of the pointings from the target ($1.9^{\circ}$). The data was taken in a series of runs with typical durations of 28\,min at a mean zenith angle of $23^{\circ}$. 

The data set was analyzed using the Hillas second moment method \citep{hillas1985} for distinguishing between $\gamma$-ray- and hadron-induced extensive air showers (EAS). 
\emph{Hard cuts} \citep{2006A&A...457..899A} were used, where at least 200 photoelectrons are required in each recorded EAS image. These cuts provide improved angular resolution\footnote{The 68\% containment radius of the PSF is 0.076$^{\circ}$.} which is crucial for this particular analysis, as well as an increased average energy threshold of 420\,GeV.
To generate 2D images, the \emph{adaptive ring background method} \citep{HESSmoriond} was used, which is a similar method to the traditional \emph{ring background method} \citep{2007A&A...466.1219B} but with the modification that the ring is allowed to grow according to the requirements in the field-of-view (FoV), and is therefore suitable when analyzing crowded regions.
A minimal inner radius of 0.6$^{\circ}$ and a ring thickness of 0.3$^{\circ}$ were used. The size of the on-region was 0.1$^{\circ}$. For the spectral analysis, the background was estimated using the \emph{reflected region method} \citep{2007A&A...466.1219B}, where the background is derived from circular off-source regions with the same angular size (0.15$^{\circ}$) and camera offset as the on-source region, located at the best-fit position of the source in question\footnote{The size of the on-source region was selected to have a source containment of roughly 90\%.}. This technique therefore minimizes systematic errors that might be introduced from an incomplete knowledge of the radial acceptance.
The statistical significances for both the images and the spectral analysis were derived from the number of off-source (background) and on-source events, following the likelihood ratio procedure in \citet{1983ApJ...272..317L}.
All presented H.E.S.S. results have been cross-checked with an alternative analysis chain using an independent calibration and $\gamma$-ray/hadron separation method \citep{2009APh....32..231D}, which give results that are consistent within statistical errors.

\subsection{Results}
\label{sec--results}
\begin{figure}[t]
\centering
  \resizebox{0.98\hsize}{!}{\includegraphics[clip=]{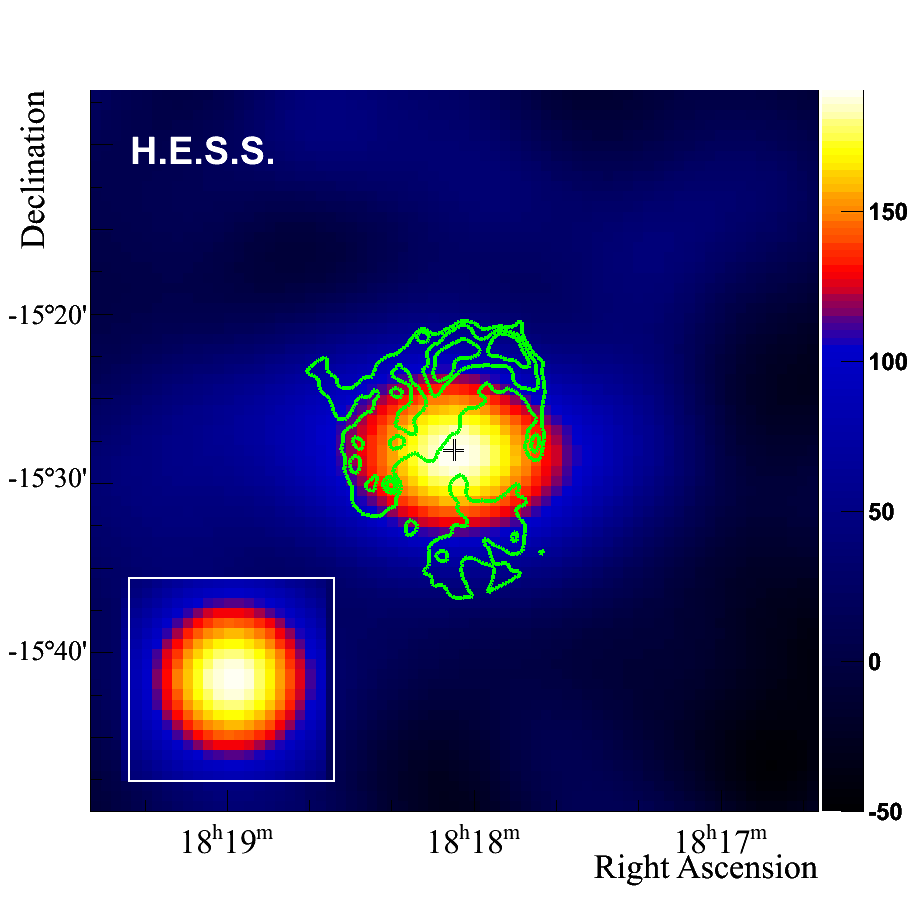}}
  \caption{The VHE $\gamma$-ray excess around the SNR \g. 
    The image has been corrected for the varying exposure across the FoV and has been smoothed with a 2D Gaussian with a width of $0.06^{\circ}$.
    The H.E.S.S. PSF for this analysis is shown in the bottom left inset.
    The color scale is chosen such that the blue-red transition occurs at roughly 4-$\sigma$ significance.
    The best-fit centroid of the $\gamma$-ray excess is indicated by a cross, the size of which
    corresponds to the sum of both statistical and systematic uncertainties. The green contours show the intensity (at 0.0175, 0.035 and 0.0525\,Jy\,beam$^{-1}$) of the radio emission from the SNR shell, based on 90-cm VLA observations \citep{1538-4357-639-1-L25}.} 
  \label{fig-tev-excess}
\end{figure}

Figure~\ref{fig-tev-excess} shows a map of the acceptance-corrected and smoothed $\gamma$-ray excess in the region around the SNR \g. A source of $\gamma$ rays from the direction of the SNR is clearly visible and detected with a significance of 8.2\,$\sigma$. Fitting the data with a 2D Gaussian model convolved with the PSF of the instrument results in a best-fit position of the source of $\alpha=18^{\mathrm{h}}18^{\mathrm{m}}4\fs8 \pm 3\fs1_{\mathrm{stat}}$ and
$\delta=-15\degr28\arcmin1\arcsec \pm 43\arcsec_{\mathrm{stat}}$\footnote{All coordinates in this work are given in J2000.0 format.} 
$(\mathrm{l} \sim 15.41^{\circ},\mathrm{b} \sim 0.16^{\circ})$,  
and the source is thus assigned the identifier \hessj. In addition to the statistical uncertainty of this fit, there is also a systematic uncertainty due to the pointing precision of the telescope array of about 20\arcsec\, \citep{Gillessen:2005wh}.
The morphology of the TeV emission is compatible with its originating in a point-like source. The 99\% confidence upper limit on the Gaussian width of the source is $0.072^{\circ}$. No significant ellipticity of the source was found.

\begin{figure}[t]
\centering
  \resizebox{0.98\hsize}{!}{\includegraphics[clip=]{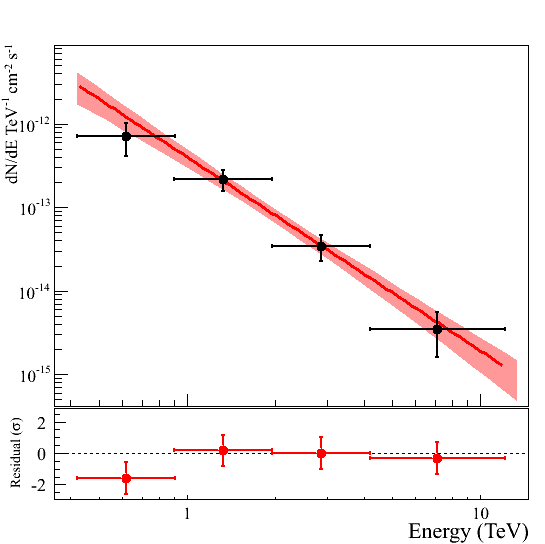}}
  \caption{The differential VHE $\gamma$-ray spectrum of \hessj. The error bars represent 1$\sigma$ statistical errors. The result of a power-law fit to the spectral data points is also shown, together with the 1\,$\sigma$ error band. The residuals between the power-law fit and the datapoints are shown in the bottom plot.} 
  \label{fig-tev-spectrum}
\end{figure}

A spectral analysis was performed and yielded 117 excess $\gamma$-ray counts in the spectral extraction region. The resulting differential spectrum is shown in Fig.\,\ref{fig-tev-spectrum} and is well fitted in the energy range 0.42\,TeV - 12.0\,TeV by a power law,

\begin{equation}
\begin{split}
\frac{\mathrm{d}N}{\mathrm{d}E} =\,\, & (0.9 \pm 0.2_{\mathrm{stat}} \pm 0.2_{\mathrm{sys}}) \times 10^{-13}\left(\frac{E}{E_{\mathrm{dec}}}\right)^{-2.3 \pm 0.3_{\mathrm{stat}} \pm 0.2_{\mathrm{sys}}} \\
&\mathrm{cm}^{-2}\mathrm{s}^{-1}\mathrm{TeV}^{-1}
\end{split}
\end{equation}
where $E_{\mathrm{dec}}=1.9$\,TeV is the decorrelation energy. This corresponds to an integral flux F$(>$0.42$\,\mathrm{TeV}$)\,=\,($0.9 \pm 0.3_{\mathrm{stat}} \pm 0.2_{\mathrm{sys}}) \times 10^{-12}$\,cm$^{-2}$\,s$^{-1}$.

\subsection{Origin of the TeV emission}
\label{sec-tev-origin}
Since the size of the shell of \g\ in radio is comparable to the size of the H.E.S.S. PSF, it is not obvious whether the observed TeV emission originates in the shell of the SNR, in an unknown source located inside the shell, or from a combination of both. Indeed, simulations show that a pure shell-type emission from \g\ convolved with the H.E.S.S. PSF results in a center-dominated source. A simulation study was therefore performed to investigate whether a shell-like origin of the TeV emission can be excluded based on the observed extension of \hessj.

\begin{figure}[t]
\centering
  \resizebox{0.98\hsize}{!}{\includegraphics[clip=]{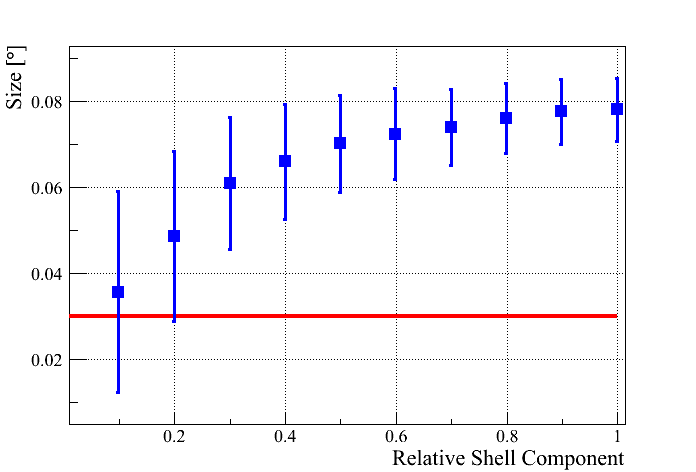}}
  \caption{The reconstructed Gaussian size for the combined TeV emission from a point-like and a shell-type source (blue datapoints). The x-axis indicates the relative contribution from the shell component to the total flux. The red line indicates the non-significant, best-fit extension of \hessj.} 
  \label{fig-tev-origin}
\end{figure}

TeV emission was simulated from two source components: a point-like and a shell-like source, the former positioned at the center of the latter. The morphology of the shell-type source was obtained by fitting a 2D projection of an elliptical 3D model of a thin shell to the SNR \g,\ as seen in 90\,cm radio data (see Sect.\,\ref{sec-radio}), and yielded a size of roughly 12\arcmin$\times$18\arcmin. 
The simulations were done using the same exposure and background maps as in the real observations of \hessj,\ and the total simulated flux was set to match the observed flux. The relative contribution of the shell component to the total flux was varied from 10\% to 100\% in steps of 10\%. For each set, 500 observations were simulated, and the width of the combined emission was derived for each simulation by fitting a 2D Gaussian model. The mean and standard deviation for each distribution of the fitted Gaussian width were then derived and are plotted in Fig.\,\ref{fig-tev-origin}. This can be compared to the best-fit extension of \hessj\ (0.03$^{\circ}$) which is indicated. This extension is not statistically significant but is nonetheless used in this particular study since it results in a more conservative result. From the observed differences in size, a scenario where a pure uniform shell-type emission is reconstructed with the Gaussian size of \hessj\ can be excluded at a significance of 6.4\,$\sigma$. Furthermore, a shell-type contribution above 45\% is excluded at 3\,$\sigma$. Models of non-uniform TeV shell emission have not been investigated, but considering that the \hess\ source is located roughly at the center of the SNR, such scenarios could probably be rejected with higher confidence than above.

\section{\xmm\ data analysis and results}
\label{sec-xmm}
To explore a PWN scenario for the newly detected compact VHE $\gamma$-ray source \hessj\ 
, the \hess\ collaboration successfully proposed to observe the region with \xmm\ to search for point-like and diffuse X-ray counterparts. 
In this section, the analysis of the European Photon Imaging Camera (EPIC) MOS \citep{2001A&A...365L..27T} and PN \citep{2001A&A...365L..18S} data from this observation (ObsID: 0691390101, PI: P. Hofverberg) is presented. 
The data were analyzed with the \xmm\ Science Analysis System (SAS) version 12.0.1, supported by tools from the FTOOLS package and XSPEC version 12.5.0 \citep{1996ASPC..101...17A} for spectral modeling. 
Some tools from the CIAO~4.4 software package were used for image processing . 

This observation was affected by some intervals of strong background flaring activity. 
To clean the data, a good time-interval (GTI) screening was employed, based on the full FoV 7--15~keV light curve provided by the standard processing chain. 
Using thresholds of 8\,cts/s for PN and 3\,cts/s for MOS, the resulting net (total) exposures are 23\,ks (30\,ks) for PN and 30\,ks (32\,ks) for MOS. 
For all spectra and images, good single and multiple events were selected: (FLAG==0), PATTERN$\leq$4 (PN), and PATTERN$\leq$12 (MOS). 

\subsection{Point-like X-ray sources}
\label{sec-xmm-point-sources}
To detect X-ray point sources, the SAS standard maximum-likelihood technique for source detection was used in several energy bands: 0.2--0.5\,keV, 0.5--1.0\,keV, 1.0--2.0\,keV, 2.0--4.5\,keV, 4.5--10.0\,keV, and 0.5--10.0\,keV. 
After merging the source lists from the three cameras and the individual energy bands, 75 unique point sources were detected in this observation. 
The observed flux of the faintest detected point source in the list is $\sim$4\ergcm{-14} (0.5--10.0\,keV, assuming a power-law spectrum with photon index $-$2), which is considered to be close to the actual detection limit considering the large sample size of 75 sources.

\begin{figure}[t]
  \resizebox{0.98\hsize}{!}{\includegraphics[clip=]{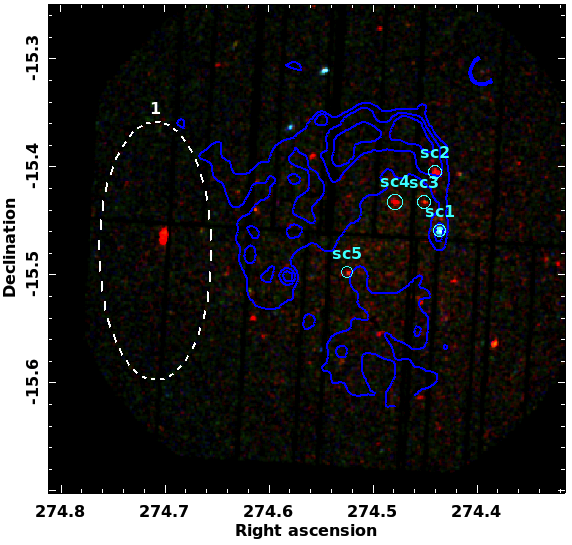}}
  \caption{Exposure-corrected and color-coded EPIC-PN counts image of the region around \hessj . 
                The energy bands are 0.5--2.0\,keV (red), 2.0--4.5\,keV (green), and 4.5--10.0\,keV (blue). 
                Over-plotted are the 90\,cm radio contours of \g\ (blue). 
                The five detected point sources compatible with the position and possible extent of \hessj\ are marked by numbered circles (cyan). 
                The dashed line (white, labeled ``1") shows the region used for the spectral analysis of the background. } 
  \label{fig-xmm-rgb-image}
\end{figure}

Because the focus of this work is to search for potential counterparts to \hessj , only the five point sources that are within a region corresponding to the location and potential size of \hessj\ are discussed below. 
These sources are labeled in Fig.~\ref{fig-xmm-rgb-image}, which shows an exposure-corrected and color-coded image created from EPIC-PN counts maps (see figure caption for details of the energy bands). 
Fortunately, the count statistics for all of these sources were sufficient to perform a spectral analysis. 
To extract source counts, a circular region was used with the 99\% containment radius of the PSF at the respective detector area, centered on the position given by the source detection algorithm. 
For each source, the background spectrum was extracted from a nearby source-free region on the same detector CCD. 
Spectra from the three EPIC cameras were fitted in parallel. 
To account for thermal and non-thermal emission, two spectral models were tested to reproduce the spectrum from each source, namely an absorbed power law and an absorbed black-body model. 
Table~\ref{tab-xmm-point-sources} summarizes the results for the five point sources. 
In each of these cases, one of the models was clearly preferred by a statistical significance of at least 3\,$\sigma$, and we only show the results for the preferred model. For some sources the fit only provided upper limits for the column density \nh . 

\begin{table*}[t]
\caption[]{X-ray point sources in the direction of \hessj}
\renewcommand{\tabcolsep}{3.5pt}
\begin{center}
\begin{tabular}{llllllllll}
\hline\hline\noalign{\smallskip}
\multicolumn{1}{l}{No.} &
\multicolumn{1}{l}{R.A.} &
\multicolumn{1}{l}{Dec.} &
\multicolumn{1}{l}{Name} &
\multicolumn{1}{l}{Counts$^{(1)}$} &
\multicolumn{1}{l}{\nh$^{(2)}$} &
\multicolumn{1}{l}{$\Gamma$ or $kT$ $^{(3)}$} &
\multicolumn{1}{l}{$F_\mathrm{X}^{(4)}$} &
\multicolumn{1}{l}{p-value$^{(5)}$} &
\multicolumn{1}{l}{Class$^{(6)}$} \\
\multicolumn{1}{l}{} &
\multicolumn{2}{c}{} &
\multicolumn{1}{l}{(XMMU)} &
\multicolumn{1}{l}{} &
\multicolumn{1}{l}{(10$^{22}$\,cm$^{-2}$)} &
\multicolumn{1}{l}{(- / keV)} &
\multicolumn{1}{l}{10$^{-14}$\,erg\,cm$^{-2}$\,s$^{-1}$} &
\multicolumn{1}{l}{} \\
\noalign{\smallskip}\hline\noalign{\smallskip}
1 &     18$^{\mathrm{h}}$17$^{\mathrm{m}}$44$^{\mathrm{s}}$ &   -15\degr 27\arcmin 33\farcs 7 & J181744-152733.7 & 1832 &   $11^{+2.7}_{-2.0}$    & \hspace{6pt}$\Gamma = -1.8^{-0.4}_{+0.3}$     &       $52.4\pm 3.9$   & 0.95 & HEXT                             \\[3pt]
2 &     18$^{\mathrm{h}}$17$^{\mathrm{m}}$46$^{\mathrm{s}}$ &    -15\degr 24\arcmin 18\farcs 9 & J181746-152418.9 & 494 &   $0.19^{+0.27}_{-0.07}$ & $kT = 0.14^{+0.02}_{-0.02}$  &        $1.8\pm 0.3$    & 0.88 & STAR                   \\[3pt]
3 &     18$^{\mathrm{h}}$17$^{\mathrm{m}}$48$^{\mathrm{s}}$ &    -15\degr 25\arcmin 57\farcs 0 & J181748-152557.0 & 128 &   $<1.9$                 & \hspace{6pt}$\Gamma = -1.9^{-0.9}_{+0.8}$ &   $1.1\pm 0.6$    & 0.39 & HGAL            \\[3pt]
4 &     18$^{\mathrm{h}}$17$^{\mathrm{m}}$55$^{\mathrm{s}}$ &    -15\degr 25\arcmin 59\farcs 0 & J181755-152559.0 & 350 &   $<0.23$                                & $kT = 0.21^{+0.03}_{-0.07}$  &        $1.3\pm 0.4$    & 0.59 & STAR                   \\[3pt]
5 &     18$^{\mathrm{h}}$18$^{\mathrm{m}}$07$^{\mathrm{s}}$ &    -15\degr 29\arcmin 55\farcs 0 & J181807-152955.0 & 152 &   $<0.95$                                &  \hspace{6pt}$\Gamma = -2.5^{-0.8}_{+0.7}$ &  $1.1\pm 0.7$  & 0.85 & HGAL             \\[3pt]
\hline\noalign{\smallskip}
\end{tabular}
\label{tab-xmm-point-sources}
\end{center}
All quoted uncertainties correspond to the 1\,$\sigma$ confidence interval; The energy range of the fit is 0.5--10.0\,keV;
$^{(1)}$Number of excess counts summed over all three EPIC cameras; 
$^{(2)}$hydrogen column density. The ``$<$"-sign indicates 99\% confidence upper limits;
$^{(3)}$power-law photon index ($\Gamma$) or black-body temperature ($kT$), depending on which model yielded the better fit; 
$^{(4)}$Observed (absorbed) energy flux (0.5--10.0\,keV) derived from the best-fit model;
$^{(5)}$fit probability;
$^{(6)}$assigned source class (see text): hard extragalactic (HEXT), hard galactic (HGAL), soft galactic (STAR);
\end{table*}

The results from the spectral analysis were used to classify the point sources to determine their most likely origin. 
Source 1 is by far the brightest source of the sample, and it features an intrinsically hard spectrum that is strongly absorbed. 
The derived value for \nh\ is a factor of $\sim$5 more than the total Galactic value toward this direction \citep[1.85\hcm{22}, see][]{1990ARAA...28..215D}. 
Therefore, this source is most likely of extragalactic origin so was classified as ``hard extragalactic (HEXT)". 
In contrast, sources 2 and 4 exhibit only very low absorption and have soft thermal spectra with temperatures of 0.1--0.2\,keV. 
Such spectra are typical of stars featuring coronal activity and/or binary interaction, which can occur during a wide variety of their evolutionary stages \citep[for a recent survey with \xmm, see][]{2013arXiv1302.6479N}. 
Therefore, these two sources were classified as ``STAR". 
The two remaining sources show intrinsically hard spectra, but their upper limits on the column density are either at the level of or below the total Galactic value, which might suggest a Galactic origin, such as cataclysmic variables, X-ray binaries, or pulsars. 
Sources 3 and 5 are thus classified as ``hard galactic (HGAL)". 

A search for periodicities in the signal of the five point sources was performed by analyzing the fast Fourier transform (FFT) power spectra derived from light curves in three energy bands: 0.5--2.0\,keV, 2.0--10.0\,keV, and 0.5--10.0\,keV. However, no significant peaks were visible above the background noise. The frame times for the \emph{full-frame} modes used in this XMM-EPIC observation are 2.6\,s for MOS and 73.4\,ms for PN, respectively, and thus there is no sensitivity for periodicities below a few times these values.

\subsection{Diffuse X-ray emission}
\label{sec-xmm-diffuse}
The second part of the \xmm\ data analysis was dedicated to the search for diffuse X-ray emission spatially coincident with the VHE $\gamma$-ray source \hessj,\ which is located in the center of the SNR \g . The procedure presented here is very similar to previous studies of diffuse X-ray emission from unidentified VHE $\gamma$-ray sources, such as HESS\,J1626$-$490 \citep{2011A&A...526A..82E} and HESS\,J1747$-$248 \citep{2010A&A...513A..66E}. 

In a first step, events around all detected X-ray point sources were removed from the EPIC event lists to facilitate analysis of only diffuse emission.
For each source, a circular exclusion region was defined, centered on the position given by the SAS source detection algorithm, with a radius equivalent to the 99\% containment radius of the PSF at the respective position in the focal plane. 
To estimate the non-X-ray background (NXB) caused by readout noise, charged particles passing the detector, etc., filter-wheel-closed datasets were used that are provided by the \xmm\ background working group \citep{2007A&A...464.1155C}. 
The same GTI-filtering criteria as for the source observation were also applied to the NXB datasets. 

\subsubsection{Maps of diffuse X-ray emission}
\label{sec-xmm-diffuse-maps}
To create maps of diffuse X-ray emission, images of X-ray counts were extracted in the two energy bands 1.0--4.0\,keV and 4.0--8.0\,keV. 
These bands were chosen to avoid the strong astrophysical background at energies below $\sim$1\,keV and a large part of the NXB that is dominant above 8--9\,keV. 
The holes in the maps resulting from the removal of point sources were refilled using the CIAO tool \textit{dmfilth} where the Poisson count statistics from annular regions around each source are used to create fake event distributions inside the excluded regions. 
Maps extracted from the NXB dataset in the same energy bands were subtracted from the source maps. 
To account for differences in exposure and in the mean level of the background count rate, the NXB maps were rescaled based on the ratio of high-energy events (10--12\,keV) between source and NXB data before subtraction \citep[a method discussed in detail by][]{2002A&A...394...77M}. 
Finally, the NXB-subtracted maps were smoothed using a Gaussian kernel with a size of 0.01$^\circ$ and divided by the exposure maps smoothed in the same manner. 
In a last step the maps from all three EPIC instruments were added for a combined image. 

\begin{figure*}[t]
  \resizebox{0.98\hsize}{!}{\includegraphics[clip=]{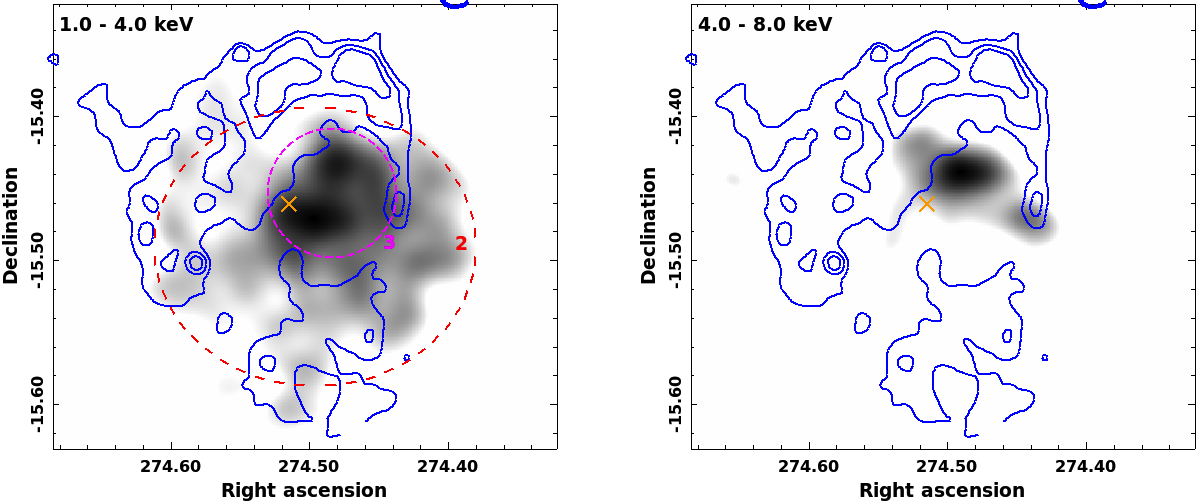}}
  \caption{Smoothed (Gaussian kernel with 0.01$^\circ$ sigma), exposure-corrected images of diffuse X-ray emission in two energy bands (as indicated in the figures). The color scale is linear between the threshold, as defined in the text, and the maximum pixel intensity. 
  The 90\,cm radio contours of \g\ are overplotted (blue). The dashed lines indicate the regions used for spectral analysis. The cross (orange) indicates the best-fit position of \hessj .} 
  \label{fig-xmm-diffuse-image}
\end{figure*}

These maps should only contain a large-scale astrophysical background component stemming from thermal emission of hot gas in the Galactic plane and potential excess emission connected to the SNR and/or the putative PWN in its center. 
The level of the diffuse astrophysical component present in these maps was estimated by the mean ($m_\mathrm{AB}$) and standard deviation ($\sigma_\mathrm{AB}$) of pixel intensities from regions outside the SNR shell, using an exclusion region centered on the SNR (R.A.: 18$^{\mathrm{h}}$18$^{\mathrm{m}}$03$^{\mathrm{s}}$, Dec.: $-$15\degr27\arcmin58\arcsec) with a radius of 8\farcm 5. 
The lower threshold (white) of the maps' color scales was set to $m_\mathrm{AB} + \sigma_\mathrm{AB}$ to significantly suppress the contribution from the astrophysical background.
Figure~\ref{fig-xmm-diffuse-image} shows the resulting maps for both energy bands. 
As suggested by these images, there is clear excess above the astrophysical background component centered inside the SNR radio shell.
Furthermore, this excess emission appears to be more compact at higher energies. 
Unfortunately, owing to the low surface brightness in the outer regions of the diffuse emission, its true extent is hard to quantify with the available data. 

\subsubsection{Spectra of the diffuse excess emission}
\label{sec-xmm-diffuse-spectrum}
The large extent and relatively low surface brightness of the diffuse excess emission makes a spectral analysis challenging, particularly concerning accurate estimation of the various background components. 
To subtract the instrumental background components, spectra extracted from the NXB dataset were used, again rescaled to match the observed count rate at high energies (see previous section). 
Owing to the energy-dependent change of the effective area with off-axis angle, one cannot simply subtract the astrophysical background using a spectrum from a region located at a very different area in the focal plane. 
Instead, one either has to correct the background spectrum for this effect \citep[see, e.g.,][]{2001A&A...365L..80A} or model the background independently \citep[as done by][]{1997ApJ...491..638K}. 
In this work the latter approach was used. 
Spectra were extracted from three regions: 

\begin{enumerate}
        \item an elliptical region outside of the radio shell and the diffuse excess to model the astrophysical background (the \emph{Background region}, shown in Fig.~\ref{fig-xmm-rgb-image}); 
        \item an elliptical region encompassing the whole X-ray excess emission (the \emph{Full region}, shown in Fig.~\ref{fig-xmm-diffuse-image});
        \item a circular region inside the SNR covering the core of enhanced X-ray emission (the \emph{Core region}, shown in Fig.~\ref{fig-xmm-diffuse-image})
\end{enumerate} 

As for the maps of diffuse X-ray emission, all detected point sources were also removed from the event lists for the spectral analysis.
The energy-dependent effective area (ARF) and energy response (RMF) files were calculated by averaging over all non-excluded pixels in the respective extraction region, assuming a flat flux distribution. 
The diffuse Galactic background spectrum was fitted with a two-temperature, non-equilibrium ionization model (2T-NEI), which is known to describe the Galactic diffuse spectrum very well \citep[for studies performed with ASCA and \chandra , see][]{1997ApJ...491..638K,2005ApJ...635..214E}. 
All model parameters except for the flux normalizations of the two thermal components were kept fixed at the best-fit values from \citep[][Table 8]{2005ApJ...635..214E}. 
The spectrum from the Background region is described very well by this model ($\chi^2$ / ndf = 1.2) with a total observed surface brightness of $(4.36\pm 0.14)\times 10^{-7}$\,erg\,cm$^{-2}$\,s$^{-1}$\,sr$^{-1}$ (0.7--10.0\,keV). 
The spectra from the full and the core regions were modeled by the sum of the best-fit 2T-NEI model (rescaled to account for the different extraction areas) and an absorbed power-law component to model the diffuse excess emission. 
Limited statistics meant that the column density of the absorbed power law was fixed at the total Galactic value in the direction of \hessj\ \citep[1.85\hcm{22}][]{1990ARAA...28..215D}.
Again, spectra from all three EPIC cameras were fit in parallel. 

\begin{table*}[tb]
\caption[]{Extraction regions and spectral fitting results for the diffuse X-ray emission}
\begin{center}
\begin{tabular}{llllllll}
\hline\hline\noalign{\smallskip}
\multicolumn{1}{l}{Reg.} &
\multicolumn{1}{l}{R.A.} &
\multicolumn{1}{l}{Dec.} &
\multicolumn{1}{l}{size$^{(1)}$} &
\multicolumn{1}{l}{area$^{(2)}$} &
\multicolumn{1}{l}{$\Gamma^{(3)}$} &
\multicolumn{1}{l}{Flux (0.5--10.0\,keV)$^{(4)}$} &
\multicolumn{1}{l}{$\chi^2$ / ndf$^{(5)}$} \\
\multicolumn{1}{l}{} &
\multicolumn{1}{l}{} &
\multicolumn{1}{l}{} &
\multicolumn{1}{l}{} &
\multicolumn{1}{l}{(10$^{-2}$\,deg$^2$)} &
\multicolumn{1}{l}{} &
\multicolumn{1}{l}{(10$^{-13}$\,erg\,cm$^{-2}$\,s$^{-1}$)} &
\multicolumn{1}{l}{} \\
\noalign{\smallskip}\hline\noalign{\smallskip}
Full            & 18$^{\mathrm{h}}$17$^{\mathrm{m}}$59$^{\mathrm{s}}$ & $-$15\degr29\arcmin2\arcsec & 6\farcm7, 5\farcm8 & 2.6      & $-3.8^{-2.6}_{+1.0}$ & 7.0$\pm$3.1    & 0.89 \\[0.1cm]
Core            & 18$^{\mathrm{h}}$17$^{\mathrm{m}}$56$^{\mathrm{s}}$ & $-$15\degr27\arcmin14\arcsec & 2\farcm7      & 0.45  & $-2.2^{-0.5}_{+0.5}$ & 3.2$\pm$0.8    & 1.1 \\[0.1cm]
\hline\noalign{\smallskip}
\end{tabular}
\label{tab-xmm-diffuse-spectra}
\end{center}
All quoted uncertainties correspond to the 1\,$\sigma$ confidence interval;
$^{(1)}$dimensions of the extraction region: semi-major and minor axes for the Full region and radius for the Core region;
$^{(2)}$effective extraction area on the detector (PN), taking excluded point sources, bad pixels/columns, and chip borders into account;
$^{(3)}$photon index of the power-law model;
$^{(4)}$unabsorbed integral flux;
$^{(5)}$value of $\chi^2$ divided by the number of degrees of freedom
\end{table*}

Table~\ref{tab-xmm-diffuse-spectra} shows the parameters of the extraction regions, as well as the best-fit results of the power-law component for the Full and the Core regions.
The quoted errors also take the uncertainties of the background model fit into account, which are, however, very small compared to the statistical uncertainties of the excess emission. 
As suggested by the different sizes of the emission in the two energy bands, the spectrum from the Core region has a harder spectrum than the Full region. 
However, this difference in photon index is only marginally significant, mostly due to large uncertainties of the spectral parameters from the Full region arising from its large size and low surface brightness resulting in a lower signal-to-noise ratio. It is worth mentioning that, given the large statistical uncertainties on the spectrum for the Full region, a thermal model is able to describe the data equally well. However, in both cases the model parameters are equally poorly constrained, and therefore the thermal results do not provide much additional information.

\section{Radio observations of \g}
\label{sec-radio}
Archival radio data from MAGPIS \citep{2005AJ....130..586W} were used to search for a counterpart to \hessj\ and the newly discovered diffuse X-ray source.
VLA observations of the region around \g\ are available at both 20\,cm and 90\,cm wavelengths, 
where the PSF (here, half-power beam-width) of the observations are $6.2^{\prime\prime}\times5.4^{\prime\prime}$ and 
$24^{\prime\prime}\times18^{\prime\prime}$, respectively. These data are sensitive to structures $<50^{\prime}$ (90\,cm) and $<18^{\prime}$ (20\,cm).
In both observations, the SNR has a clear shell structure and appears largely void of emission from the central regions (see Fig.~\ref{fig-tev-excess}). Therefore, since there is no obvious radio counterpart to the X-ray and TeV source, an upper limit on the flux density from a region corresponding to the Core region (as defined in Sect.~\ref{sec-xmm}) was derived. An upper limit on the flux density from the Full region was obtained by scaling the result from the Core region with the ratio of the area between the two regions. This procedure minimizes the contamination of the emission from the shell, while still being conservative, since the most luminous part of the X-ray source is contained inside the Core region.
The images were first convolved for a common beam size of $25\arcsec \times 25\arcsec$, and
point sources\footnote{From SIMBAD, http://simbad.u-strasbg.fr/simbad} were removed by masking areas 1.5 times the beam size. 
The RMS noise was then derived from each (convolved) image and was found to be $\sigma_{90} = 20$\,mJy\,beam$^{-1}$ and
$\sigma_{20} = 4$\,mJy\,beam$^{-1}$, respectively. 
After integrating within the Core and the Full regions, the 3\,$\sigma$ flux upper limits given in Table~\ref{tab:radioflux} were obtained. 
Uncertainties in the flux density include contributions from i) image noise, ii) calibration, and iii) wrong zero levels, 
as suggested by \citet{2003A&A...406..579K}.

\begin{table}[h]
\centering
\begin{tabular}{cccc}
\hline\hline
Region & Flux (90~cm) & Flux (20~cm) \\
 & (Jy) & (Jy)\\
\hline
Full & <4.8 & <9.6 \\
Core & <0.9 & <1.8 \\
\hline
\end{tabular}
\caption{The 3\,$\sigma$ flux upper limits from the Core and the Full regions, respectively.}
\label{tab:radioflux}
\end{table}

\section{Discussion}
\label{sec-discussion}

\subsection{A PWN inside \g\ detected in X-rays and VHE $\gamma$ rays}
The observed X-ray emission exhibits a relatively luminous core component completely contained within the boundaries of the radio shell of \g. It is well fit by a hard spectrum power law, which is a clear indication of non-thermal radiation. Surrounding the core component and roughly centered on it, a diffuse X-ray component
is observed that fades with distance from the core component. 
The diffuse emission completely fills the inside of the SNR and partly overlaps with the weaker regions of the radio shell. However, there is no overlap with the northern part of the SNR where the radio emission is strongest.
It is therefore very likely that the observed X-ray emission stems from a previously unknown PWN, in which a centrally located pulsar is driving a wind of relativistic leptons that radiates in X-rays via synchrotron mechanism. The discovery of a morphologically compatible TeV counterpart strengthens this case further, since TeV $\gamma$ rays can be produced by the same population of leptons through IC scattering on (predominately) cosmic microwave background (CMB) photons. 
The X-ray and VHE $\gamma$-ray observations of \g, presented in this paper, therefore show evidence of a PWN located inside the shell of the SNR and thus establish this object as a \emph{composite} SNR.

X-ray PWNe are generally brighter towards the pulsar thanks to the cooling of the electrons as they propagate outward from the acceleration site. This is also true for \g, with the difference that the core appears more like a distinct, well-defined region superposed on a much fainter diffuse background. Another possibility is then that the Core region represents the termination shock of the pulsar and that its somewhat elongated morphology, observed at high energies, is caused by an anisotropic outflow from the pulsar. 
However, the extension of the termination shock in PWNe is normally in the range of 0.03-0.3~pc \citep{2009ASSL..357..451D}, which even for a small distance to \g\ of 1~kpc, translates to $<$$1^{\prime}$ and would thus appear point like to \xmm, unlike what is observed in the high-energy X-ray image. The compact feature seen in X-rays is therefore most likely dominated by synchrotron emission from the PWN.

Another possibility is then that the X-ray emission from the Core region is in fact the complete PWN, while the diffuse emission is simply thermal emission from the shell. However, the lack of emission from the bright parts of the radio shell makes it unlikely that this is a dominant part of the emission. It is therefore assumed in the following that the complete observed X-ray emission originates in the cooled wind of the PWN with the caveats noted that part of the emission can come from the pulsar, the termination shock, and from thermal emission from shock-heated material.

Since the extension of the PWN in X-rays and VHE $\gamma$ rays is comparable, the broad-band emission can be assumed to originate in a single population of particles. A one-zone model can therefore be used to model the emission in the context of a PWN scenario. A purely static leptonic model was considered here, where the PWN hosts a population of leptons with a power-law spectrum and an exponential cutoff, inside a region with a uniformly distributed magnetic field. The synchrotron and IC emission from this lepton population were then calculated, only taking IC scattering on CMB photons into account.

\begin{figure}[t]
\centering
  \resizebox{0.98\hsize}{!}{\includegraphics[clip=]{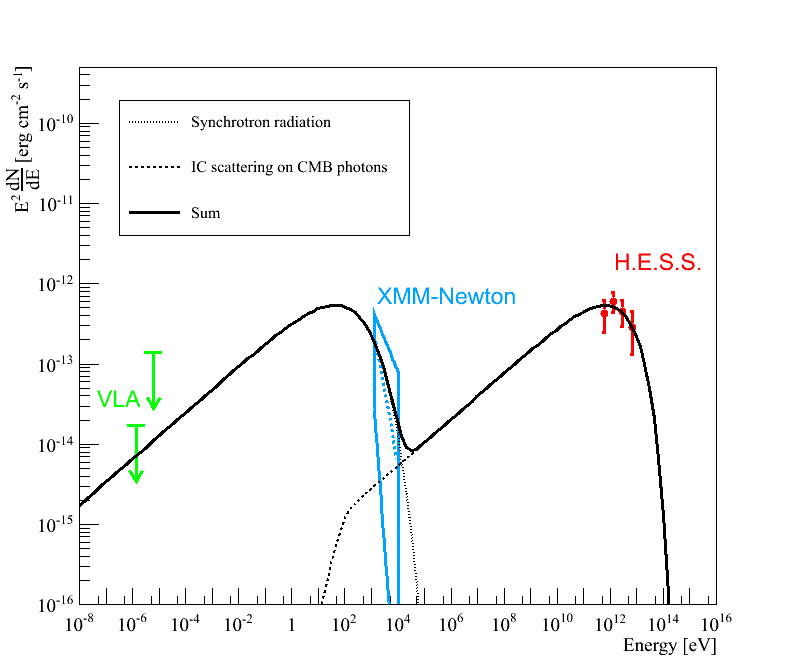}}
  \caption{The spectral energy distribution of the PWN inside \g\ from radio to VHE $\gamma$ rays. The best-fit X-ray spectrum is shown with the 1\,$\sigma$ error band of the fit. Only statistical errors are shown. Overlaid on the datapoints is the best-fit one-zone leptonic model of the PWN. } 
  \label{fig-sed-model}
\end{figure}

Figure~\ref{fig-sed-model} shows the broad-band spectral energy distribution (SED) of the newly discovered PWN, together with the best-fit one-zone model to the radio, X-ray, and $\gamma$-ray data overlaid\footnote{The de-absorbed X-ray data points depend on the parameters of the absorption model.}. The radio and X-ray flux points were derived from the Full region. The free parameters of the model are the total energy in relativistic leptons, the spectral index and exponential cutoff energy of the lepton population, and the average magnetic field. 
The 68\,\% confidence intervals for the fitted model parameters are given in Table~\ref{tab-sed-param}.
Although the model is very simple, it describes the data well with reasonable physical properties and predicts a low average magnetic field inside the PWN. To better constrain the total energy in the system and the spectral index of the lepton distribution, a detection of the PWN in the radio regime is necessary. A model with a low energy cutoff in the parent lepton distribution, which is expected in the PWN paradigm \citep[see e.g.][]{2006MNRAS.367..937W}, has not been considered here.

\begin{table}[h]
\centering
\begin{tabular}{cc}
\hline\hline\noalign{\smallskip}
Parameter & Value \\
\noalign{\smallskip}\hline\noalign{\smallskip}
Magnetic field ($\mu$G) & 1.5\,--\,4.8 \\
Spectral index $\left(\frac{\mathrm{d}N}{\mathrm{d}E}\sim E^{-p}\right)$ & < 2.6 \\
Exp. cutoff (TeV) & 10\,--\,80 \\
Total energy (erg)      & < $3\times10^{49}$ \\
\noalign{\smallskip}\hline\noalign{\smallskip}
\end{tabular}
\caption{The 68\,\% confidence intervals for the parameters of the one-zone model in Fig.~\ref{fig-sed-model}. One-sided limits are given for the spectral index and total energy since their $\chi^2$ contours are highly non-parabolic for lower values than the best-fit value. The total energy is given for a distance of 4.8~kpc.}
\label{tab-sed-param}
\end{table}

\subsection{A hadronic scenario for the TeV emission?}
\citet{castelletti2013} has recently reported the discovery of a molecular cloud in spatial coincidence with the northern part of the SNR shell. The molecular cloud, about 5$'$ in size, consists of two bright clumps with masses and densities on the order of $10^3$\,M$_{\sun}$ and $10^3$\,cm$^{-3}$, respectively. From the agreement of the kinematical distances between the SNR and the cloud and from tentative evidence showing that the northern cloud is being disrupted by a strong shock, it was argued that the SNRs interact with the molecular cloud. The authors furthermore show that the observed TeV flux \citep{2011ICRC....7..247H} can be produced by the decay of $\pi^0$ mesons created in the interaction between accelerated protons/nuclei and the molecular cloud, assuming a conversion of a few percent of the supernova explosion energy into relativistic protons.

The molecular cloud extends well outside the shell of the northern boundary of the SNR, and it is therefore clear that the TeV emission cannot originate in the complete cloud. The southern clump is closer to the center of the SNR, hence also to the TeV source, and it cannot be ruled out with high confidence that they overlap. A hadronic origin of the TeV emission is therefore possible, although the lack of emission from the direction of the northern part of the cloud disfavors this scenario.

Also in a hadronic scenario as proposed above, a certain amount of X-ray emission is expected from bremsstrahlung and synchrotron emission from secondary electrons \citep{Gabici01072009}. This can, however, not explain the observed X-ray emission observed from \g\ since there is a clear separation between the molecular cloud and the X-ray emission region. 

A scenario with a common origin of the X-ray and TeV emission from a PWN located inside the shell of \g\ is therefore preferred.

\subsection{Age and evolution of the SNR \g\ and its PWN}
Little is known about the age of the SNR \g\ and its putative pulsar. \citet{castelletti2013} placed the SNR at a distance of ($4.8\pm1.0$)\,kpc. At this distance, the angular size of the shell of \g\ (12$'$$\times$18$'$) corresponds to a physical size of (17$\times$15)\,pc. From the average physical size of the SNR and assumptions on the ambient medium density and the supernova explosion energy, the Sedov-Taylor model of SNR evolution from \cite{blondin1998} can be used to derive a rough estimate of the age of the SNR. Since the ambient density around \g\ is unknown, a value of 1\,cm$^{-3}$ was adopted, which is valid for a large number of SNRs in the galactic plane. A typical supernova explosion energy of $10^{51}$\,erg was assumed. The model then predicts an age of about 2500\,years, placing this SNR in the Sedov phase.

A scenario with a rather young, unevolved system is also supported by the morphology of the SNR and the PWN as seen in radio, X-ray, and TeV energies, assuming the latter is indeed the high-energy counterpart of the PWN. With a X-ray and TeV PWN centered inside the SNR shell, and a roughly symmetric diffuse X-ray component surrounding the inner core, it is unlikely that the reverse shock, which normally hits the expanding PWN after $\sim$2000~yrs \citep{2001A&A...380..309V}, has had time yet to significantly alter the system.

The age of the putative pulsar can be estimated using the relationship between the $\gamma$-ray to X-ray flux ratio and the characteristic age of the pulsar proposed by \citet{2009ApJ...694...12M} \citep[and further developed by][]{2013arXiv1305.2552K}. This yields a characteristic age of the pulsar of 17\,kyr. Interestingly, this suggests an evolved system, in contradiction to the above estimations. However, there are several examples of sources where the true age is significantly less than the characteristic age of the pulsar where the discrepancy is attributed to a fairly constant pulsar spin-down \citep{2003ApJ...588..992R,2010ApJ...716..663R}.

\subsection{Non-detection of a pulsar}
Both the hard galactic (HGAL, see Table~\ref{tab-xmm-point-sources}) X-ray point sources qualify as potential candidate pulsars powering the observed PWN.
However, both these sources are located at significant offsets from the peak of the diffuse X-ray emission, which is generally expected to be close to the location of the pulsar, particularly in younger systems. It therefore seems unlikely that these correspond to the pulsar powering this synchrotron nebula. 

The non-thermal X-ray emission from a PWN is usually a factor of $\sim$10 greater than the X-ray emission from its pulsar \citep{2008ApJ...682.1166L}. The expected X-ray flux from the putative pulsar associated to \hessj\ should therefore be on the order of $(10^{-14}-10^{-13})$\,erg\,cm$^{-2}$\,s$^{-1}$ (0.5--10.0\,keV), which is considerably higher than the point-source sensitivity of the current \xmm\ observation. A non-detection of a pulsar candidate in this observation would therefore be surprising. 
However, there are some cases where predominantly thermal X-ray radiation has been detected from pulsars powering young PWNe with black-body temperatures of $\sim$0.1\,keV \citep[see, e.g.,][]{2004ApJ...612..389H}. 
If this was the case for \hessj , such a source might be challenging to detect, given the high expected absorption column density due to the SNRs large distance. 

The VLA observations of \g\ show no signs of a pulsar either, but in this case it is not surprising given the sensitivity of the observations. A typical young radio pulsar located at 4.8\,kpc with an assumed luminosity of $\mathrm{L}_{20\,\mathrm{cm}}\sim$2.4\,$\mathrm{mJy}\cdot\mathrm{kpc}^2$  \citep[like the median for low-luminosity young rotation powered pulsars as estimated by][]{2006ApJ...637..456C} would have a flux density of $\mathrm{S}_{20\,\mathrm{cm}}\sim 0.1~\mathrm{mJy}$, considerably lower than the sensitivity of current VLA data. Furthermore, beaming effects could also make the pulsar virtually undetectable in radio.

\section{Conclusions}
\label{sec-conclusions}
A new VHE $\gamma$-ray source, \hessj, has been discovered toward the direction of the SNR \g\ with a flux above 0.42\,TeV of $(0.9 \pm 0.3_{\mathrm{stat}} \pm 0.2_{\mathrm{sys}}) \times 10^{-12}$\,cm$^{-2}$\,s$^{-1}$. 
The centroid of the source is at, or near, the center of the SNR, and a pure shell-type origin of the emission can be ruled out with high confidence from the observed size of the source. The VHE $\gamma$-ray emission therefore likely originates in a previously unknown source located inside the shell of \g.

Follow-up observations of \hessj\ with \xmm\ have revealed a diffuse source of X-rays coincident with the TeV emission. The X-ray emission exhibits a two-component morphology with a hard-spectrum core component surrounded by a fainter halo that fills the SNR cavity completely. The energy flux of the total diffuse X-ray emission is ($7.0\pm 3.1$)\ergcm{-13} (0.5--10.0\,keV). The observed VHE $\gamma$-ray emission is morphologically compatible with the extended X-ray source.

The morphological and spectral characteristics of the X-ray emission provide strong evidence of a PWN located inside the shell of \g. The detection of a morphologically compatible TeV counterpart to the X-ray PWN further strengthens this scenario. These observations therefore classify this object as a \emph{composite} SNR. Only a handful of such sources have been detected at TeV energies, and \g\ is the first case where the composite nature has been discovered first and classified on the basis of VHE $\gamma$-ray observations.
A one-zone leptonic model describes the broad-band emission well, predicting a low average magnetic field inside the PWN.
Two hard-spectrum point sources discovered in the \xmm\ observation qualify as potential candidate pulsars powering the observed PWN. However, neither of them constitutes a strong candidate since they are both offset relative to the peak X-ray emission. No pulsar candidates have been discovered in archival radio observations of the SNR.

\citet{castelletti2013} recently discovered a molecular cloud in spatial coincidence with the northern part of the SNR shell and then suggested a hadronic origin of the TeV emission. Although this scenario cannot be excluded, the so far tentative evidence for an interaction between the MC and the shell, the poor morphological match between the MC and the TeV source, and the need to invoke an additional mechanism to explain the X-ray emission makes a hadronic scenario unlikely.

Future deep radio observations are promising to confirm that a powerful pulsar is located at the center \g . 
Also, searches for a counterpart to the X-ray synchrotron nebula in radio would allow the injection spectrum of the underlying lepton population to be better constrained. 
Furthermore, extending the VHE $\gamma$-ray coverage into lower energies would probably facilitate the detection of the IC peak, which is another crucial component for further constraining the parameters of the lepton population. Thus, future observations with \hess\,II and CTA are highly desirable. 

\begin{acknowledgements}
  The support of the Namibian authorities and of the University of Namibia
  in facilitating the construction and operation of H.E.S.S. is gratefully
  acknowledged, as is the support by the German Ministry for Education and
  Research (BMBF), the Max Planck Society, the German Research Foundation (DFG), 
  the French Ministry for Research,
  the CNRS-IN2P3 and the Astroparticle Interdisciplinary Programme of the
  CNRS, the U.K. Science and Technology Facilities Council (STFC),
  the IPNP of the Charles University, the Czech Science Foundation, the Polish 
  Ministry of Science and  Higher Education, the South African Department of
  Science and Technology and National Research Foundation, and the
  University of Namibia. We appreciate the excellent work of the technical
  support staff in Berlin, Durham, Hamburg, Heidelberg, Palaiseau, Paris,
  Saclay, and in Namibia in the construction and operation of the
  equipment. This research has made use of software provided by the Chandra X-ray Center (CXC) in the application package CIAO. 
\end{acknowledgements}

\bibliographystyle{aa}
\bibliography{citations}

\end{document}